\documentclass[11pt,english]{article}
\usepackage[T1]{fontenc}
\usepackage[latin9]{inputenc}
\usepackage{amssymb}
\usepackage{esint}

\makeatletter
\usepackage{amssymb}
\usepackage{latexsym}
\usepackage{epsfig}

\newcommand{\be}{\begin{equation}}
\newcommand{\ee}{\end{equation}}

\makeatother

\usepackage{babel}
\begin{document}
{}~ \hfill\vbox{\hbox{CTP-SCU/2015001}}\break
\vskip 3.0cm
\centerline{\Large \bf Non-commutativity from the double sigma model}
\vspace*{10.0ex}
\centerline{\large Dimitri Polyakov, Peng Wang, Houwen Wu and Haitang Yang}
\vspace*{7.0ex}
\vspace*{4.0ex}
\centerline{\large \it Center for theoretical physics}
\centerline{\large \it Sichuan University}
\centerline{\large \it Chengdu, 610064, China} \vspace*{1.0ex}
\centerline{polyakov, pengw, hyanga@scu.edu.cn, iverwu@stu.scu.edu.cn}
\vspace*{10.0ex}
\centerline{\bf Abstract} \bigskip \smallskip
We show how non-commutativity arises from commutativity in the double sigma model. We demonstrate that this model is intrinsically non-commutative by calculating the propagators.  In the simplest phase configuration, there are two dual copies of commutative theories. In general rotated frames, one gets a non-commutative theory and a commutative partner. Thus a non-vanishing $B$ also leads to a commutative theory. Our results imply that $O\left(D,D\right)$ symmetry unifies not only the big and small torus physics, but also the commutative and non-commutative theories. The physical interpretations of the metric and other parameters in the double sigma model are completely dictated by the boundary conditions. The open-closed relation is also an $O(D,D)$ rotation and naturally leads to the Seiberg-Witten map. Moreover, after applying a second dual rotation, we identify the description parameter in the Seiberg-Witten map  as an $O(D,D)$ group parameter and all theories are non-commutative under this composite rotation.  As a bonus, the propagators of general frames in double sigma model for open string are also presented.

\vfill 
\eject
\baselineskip=16pt
\vspace*{10.0ex}
\tableofcontents

\section{Introduction}

Symmetries and dualities play a central role in modern physics. There
are some evidences that various dualities in string theory may emerge
from symmetry breaking in some select backgrounds. In string theory,
there are three well known dualities: T-duality relates large and
small distance physics; S-duality connects weak and strong couplings;
and U-duality unifies the T-duality and S-duality. M-theory is dictated
by these three types of dualities. There is another significant equivalence
in string theory, the Seiberg-Witten map \cite{Seiberg:1999vs} which
seems to disconnect from known dualities or symmetries. The Seiberg-Witten
map relies on the non-commutativity from string theory and the open-closed
relation. In the language of string theory, commutativity is a natural
description for closed string and non-commutativity is natural for
open string. The Seiberg-Witten map is a map between commutative variables
$\left(\hat{g}_{s},\hat{g}{}_{ij},\hat{B}{}_{ij}\right)$ of closed
strings and non-commutative variables $\left(g_{s},g_{ij},B_{ij}\right)$
of open strings. The Dirac-Born-Infeld (DBI) action, obtained from
calculating the one-loop beta function of worldsheet action, is originally
expressed with commutative variables $\left(\hat{g}_{s},\hat{g}{}_{ij},\hat{B}{}_{ij}\right)$.
Utilizing the Seiberg-Witten map, one can derive a non-commutative
version of the DBI action. Therefore, the Seiberg-Witten map unifies
the non-commutative gauge theories and commutative gauge theories.

In this work, we present solid evidences to show that the Seiberg-Witten
map can be described by a pure coordinate transformation of $O(D,D)$
group. The relationship between non-commutative property and T-duality
was first proposed by Connes, Douglas and Schwarz \cite{Connes:1997cr},
where they showed that the Matrix model on tori with an anti-symmetric
field produces non-commutativity in a natural way. Non-commutativities
from open string ending on D-branes are first addressed in \cite{Chu:1998qz}.
In \cite{Maharana:2000fc} , Maharana and Pal made a nice try to construct
a relationship between $O\left(D,D\right)$ transformations and the
Seiberg-Witten map. They introduced T-dual coordinates in an ad hoc
way and found some links between $O\left(D,D\right)$ and the Seiberg-Witten
map. But since their setup is not a consistent $O\left(D,D\right)$
covariant theory, the underlying connection of $O\left(D,D\right)$
and the Seiberg-Witten map was not revealed. 

The $O\left(D,D\right)$ symmetry is a continuous symmetry for non-compact
background, where $D$ is a number of spacetime dimensions. If we
compactify $d=D-n$ dimensions, the continuous $O\left(D,D\right)$
breaks into an $O\left(n,n\right)\times O\left(d,d;\mathbb{Z}\right)$
group. The $O\left(d,d;\mathbb{Z}\right)$ group is known as T-duality
group in the compactified background. Moreover, it is well known that
the solution space of the closed string low energy effective action
possesses $O(n,n)$ symmetry for certain backgrounds. These features
makes it tempting to construct theories having $O(D,D)$ invariance
at the very beginning. The double sigma model was proposed by Tseytlin
\cite{Tseytlin:1990nb} and developed in \cite{Maharana:1992my,Schwarz:1993vs}
to fulfill this purpose. Recent developments of double field theory
\cite{Siegel:1993xq,Hull:2009mi,Duff:1989tf,Berman:2007xn} have the
similar inspirations. In the double sigma action, the $O(D,D)$ invariance
is manifested by introducing another set of target space coordinates.
In the same pattern of Polyakov action, open string and closed string
share the same double sigma action. Whether the theory represents
open or closed string is determined by the boundary conditions. In
this paper, we will address the open string scenario. Surprisingly,
after calculating the propagators, we find that the double sigma model
intrinsically has both commutativity and non-commutativity. Starting
from the simplest phase configuration, there are two commutative theories,
either of which can represent our ordinary picture. However, the hidden
mixed propagators between the dual theories exhibit non-commutative
property, which implies that the ordinary coordinate $X$ is non-commutative
to its dual $\tilde{X}$. With this property, it is not hard to see
that after an $O(D,D)$ rotation, a non-commutative theory emerges.
Our calculation shows that this non-commutative theory is completely
identical to that in \cite{Seiberg:1999vs}, \cite{Fradkin85,Callan87}
. It turns out in general phase frame, besides the non-commutative
theory, we have another commutative companion even if the non-commutative
parameter is nonvanishing. The Neven-Schwartz $B$ field is simply
a group parameter in this theory. Another very interesting observation
is that, unlike the situation in the Polyakov action, the physical
interpretations of the metric and other parameters in the double sigma
action are completely dictated by the boundary conditions. The parameters
between closed and open strings are linked by an $O(D,D)$ transformation.
With all these results in mind, we conclude that the double sigma
action unifies both commutativity and non-commutativity and Seiberg-Witten
map can be described by a subset of $O(D,D)$ group. It is quite inspiring
to notice that $O\left(D,D\right)$ group not only includes dual theories
of small and large tori, but also unifies the commutative and non-commutative
theories. The unification of the commutative and non-commutative theories
is realized by the intrinsic non-commutative property of the action,
observation of commutativity or non-commutativity depends on which
frame is taken. 

It is remarkable that after performing another dual rotation with
a group parameter $C$, we find that the description parameter $\Phi$
in the Seiberg-Witten map is not necessary any more and can be removed.
For fixed closed string parameters, the corresponding open string
parameters are not unique. The arbitrariness of the non-commutative
parameter $\theta$ is accounted by the the alterable open string
metric $g$, which in turn derived from the arbitrariness of the group
parameter $C$. Therefore, $C$ replaces the role of $\Phi$ to make
the non-commutative theory is optional for different metric and {*}
product parameter $\theta$. Another feature of the composite rotation
is that both the ordinary theory and its dual become non-commutative. 

The reminder of this paper is outlined as follows. In section 2, we
briefly review the non-commutativity in string theory and the Seiberg-Witten
map. By using the double sigma model, we derive the non-commutative
theory from a commutative one through $O\left(D,D\right)$ coordinate
transformations in section 3. We address the general description under
a composite rotation in section 4.

\section{Non-commutativity in string theory and the Seiberg-Witten map}

This short review is based on Seiberg and Witten's work \cite{Seiberg:1999vs},
and Mukhi's talk \cite{Sunil Mukhi}. We begin with Euclidean worldsheet
action

\begin{equation}
S=\frac{1}{2}\int_{\Sigma}\left(g_{ij}\partial_{\alpha}X^{i}\partial^{\alpha}X^{j}-i\epsilon^{\alpha\beta}B_{ij}\partial_{\alpha}X^{i}\partial_{\beta}X^{j}\right),
\end{equation}

\noindent where $\epsilon$ is an anti-symmetric matrix, $g$ is the
spacetime metric. The worldsheet equation of motion is given by

\begin{equation}
g_{ij}\partial_{\alpha}\partial^{\alpha}X^{j}=0.
\end{equation}

\noindent The open string boundary condition is 

\begin{equation}
\left.g_{ij}\partial_{\sigma}X^{j}+iB_{ij}\partial_{\tau}X^{i}\right|_{\partial\Sigma}=0.
\end{equation}

\noindent To calculate the correlation function, we map the disc to
the upper half $z$ plane by $z=\tau+i\sigma$. The propagator is

\begin{eqnarray}
\left\langle X^{i}\left(z,\bar{z}\right)X^{j}\left(z^{\prime},\bar{z}^{\prime}\right)\right\rangle  & = & -\alpha^{\prime}\left[g^{ij}\log\left|z-z^{\prime}\right|-g^{ij}\log\left|z-\bar{z}^{\prime}\right|\right.\nonumber \\
 &  & \left.+G^{ij}\log\left|z-\bar{z}^{\prime}\right|^{2}+\frac{1}{2\pi\alpha^{\prime}}\theta^{ij}\log\frac{z-\bar{z}^{\prime}}{\bar{z}-z^{\prime}}+D^{ij}\right],
\end{eqnarray}
where

\begin{equation}
G_{ij}=\left(g-Bg^{-1}B\right)_{ij},\qquad\theta^{ij}=-\left(\frac{1}{g-B}B\frac{1}{g+B}\right)^{ij},
\end{equation}

\noindent and can be grouped as

\begin{equation}
\frac{1}{G}+\theta=\frac{1}{g+B}.
\end{equation}

\noindent Now, consider the propagator on the boundary and one gets

\begin{equation}
\left\langle X^{i}\left(\tau\right)X^{j}\left(\tau^{\prime}\right)\right\rangle =-\alpha^{\prime}\left[2G^{ij}\log\left|\tau-\tau^{\prime}\right|+\frac{i}{2}\theta^{ij}\varepsilon\left(\tau-\tau^{\prime}\right)\right],
\end{equation}

\noindent where $\varepsilon\left(\tau-\tau^{\prime}\right)=+1$ when
$\tau-\tau^{\prime}>0$, and $\varepsilon\left(\tau-\tau^{\prime}\right)=-1$
when $\tau-\tau^{\prime}<0$. Then, $G^{ij}$ is effectively identified
as the open string metric. The antisymmetric quantity $\theta^{ij}$
introduces non-commutativity into the theory. To see the map between
commutative and non-commutative gauge theories, we turn to the DBI
action 

\begin{equation}
S_{\mathrm{DBI}}=\frac{1}{g_{s}}\int d^{D}x\sqrt{\det\left(g+B+F\right)},
\end{equation}

\noindent where $g_{s}$ is the closed string coupling and $F_{ij}=\partial_{i}A_{j}-\partial_{j}A_{i}$
is the gauge field strength. For simplicity, the field strength $F$
is assumed to be constant. Expanding this action in order of $F$,
one gets the Maxwell equation where $S\sim\int F_{ij}F^{ij}$. Applying
the open-closed string relation between the closed string side $\left(g,B\right)$
and the open string side $\left(G,\theta\right)$, the DBI action
can be rewritten in the terms of $\left(G,\theta\right)$ as

\begin{equation}
S_{\mathrm{DBI}}=\frac{1}{g_{s}}\int d^{D}x\frac{\sqrt{\det\left(1+\theta F\right)}}{\sqrt{\det\left(1+\theta G\right)}}\sqrt{\det\left(G+F\frac{1}{1+\theta F}\right)}.
\end{equation}
After redefining 

\begin{equation}
\hat{F}\equiv F\frac{1}{1+\theta F},\qquad G_{s}\equiv g_{s}\sqrt{\det\left(1+\theta G\right)}.
\end{equation}

\noindent the DBI action becomes

\begin{equation}
S_{\mathrm{DBI}}=\frac{1}{G_{s}}\int d^{D}x\frac{1}{\sqrt{\det\left(1-\theta\hat{F}\right)}}\sqrt{\det\left(G+\hat{F}\right)}.
\end{equation}

\noindent The coupling $G_{s}$ can be seen as a new string coupling,
and $B$ field disappeared. Since the prefactor $\sqrt{\det\left(1-\theta\hat{F}\right)}$
can be canceled by open Wilson line or understood as a nontrivial
Jacobian factor from a local coordinate transformation eliminating
$U(1)$ gauge fields \cite{Yang:2004vd} and does not affect our discussion,
the action is simplified as

\begin{equation}
S_{\mathrm{DBI}}=\frac{1}{G_{s}}\int d^{D}x\sqrt{\det\left(G+\hat{F}\right)}.
\end{equation}

\noindent To understand the physical implication of $\hat{F}$, we
expand it in order of $\theta$, 

\begin{equation}
\hat{F}_{ij}=F_{ij}-F_{ik}\theta^{kl}F_{lj}+\mathcal{O}\left(\theta^{2}\right).
\end{equation}

\noindent Since $F_{ij}=\partial_{i}A_{j}-\partial_{j}A_{i}$, if
we redefine a new gauge potential $\hat{A}_{i}$ as

\begin{equation}
\hat{A}_{i}\equiv A_{i}-\theta^{kl}\left(A_{k}\partial_{l}A_{i}+\frac{1}{2}A_{k}\partial_{i}A_{l}\right)+\mathcal{O}\left(\theta^{2}\right).
\end{equation}

\noindent The $\hat{F}$ can be rewritten as

\begin{eqnarray}
\hat{F}_{ij} & = & \partial_{i}\hat{A}_{j}-\partial_{j}\hat{A}_{i}+\theta^{kl}\partial_{k}\hat{A}_{i}\partial_{l}\hat{A}_{j}\nonumber \\
 & = & \partial_{i}\hat{A}_{j}-\partial_{j}\hat{A}_{i}+\left\{ \hat{A}_{i},\hat{A}_{j}\right\} ,
\end{eqnarray}

\noindent where the Poisson bracket $\left\{ \hat{A}_{i},\hat{A}_{j}\right\} $
can be lifted to a non-commutative commutator as $-i\left[\hat{A}_{i},\hat{A}_{j}\right]_{*}$
through the Moyal-Weyl product as follows

\begin{equation}
\left\{ f,g\right\} \rightarrow-i\left(f*g-g*f\right),
\end{equation}

\noindent with 

\begin{equation}
f\left(x\right)*g\left(x\right)=\left.e^{\frac{i}{2}\theta^{ij}\frac{\partial}{\partial x^{i}}\frac{\partial}{\partial y^{j}}}f\left(x\right)g\left(y\right)\right|_{x=y}.
\end{equation}

\noindent Therefore, we have two copies of DBI actions, one is commutative

\begin{equation}
S_{\mathrm{DBI}}=\frac{1}{g_{s}}\int d^{D}x\sqrt{\det\left(g+B+F\right)},
\end{equation}

\noindent the other is non-commutative

\begin{equation}
S_{\mathrm{DBI}}=\frac{1}{G_{s}}\int d^{D}x\sqrt{\det\left(G+\hat{F}\right)}.
\end{equation}

\noindent where $\hat{F}_{ij}=\partial_{i}\hat{A}_{j}-\partial_{j}\hat{A}_{i}-i\left[\hat{A}_{i},\hat{A}_{j}\right]_{*}$.
The relations between these two DBI actions, or relations between
commutative gauge theories and non-commutative gauge theories are

\begin{eqnarray}
\hat{F} & = & F\frac{1}{1+\theta F},\nonumber \\
\hat{A}_{i} & = & A_{i}-\theta^{kl}\left(A_{k}\partial_{l}A_{i}+\frac{1}{2}A_{k}\partial_{i}A_{l}\right)+\mathcal{O}\left(\theta^{2}\right).
\end{eqnarray}

\section{$O(D,D)$ rotations of the propagators}

To understand how the non-commutativity arises from $O(D,D)$ symmetry,
we start from the simplest phase configuration of the double sigma
model. The action with Lorentz signature is given as follows

\noindent 
\begin{equation}
S=-\frac{1}{4\pi\alpha'}\int_{\Sigma}\left(-\partial_{1}X^{M}h_{MN}\partial_{1}X^{N}+\partial_{1}X^{M}\eta_{MN}\partial_{0}X^{N}\right),\label{eq:Double Sigma Action}
\end{equation}

\noindent where $\partial_{0}=\partial_{\tau}$, $\partial_{1}=\partial_{\sigma}$
and

\noindent 
\begin{equation}
h_{MN}=\left(\begin{array}{cc}
g_{ij} & 0\\
0 & g^{ij}
\end{array}\right),\qquad\eta_{MN}=\left(\begin{array}{cc}
0 & 1\\
1 & 0
\end{array}\right),\qquad X^{M}=\left(\begin{array}{c}
X^{i}\\
\tilde{X}_{i}
\end{array}\right).
\end{equation}

\noindent In the double sigma action, the $D$ dimensional target
space coordinate $X^{i}$ is doubled to $2D$ dimensional $X^{M}\left(X^{i},\tilde{X}_{i}\right)$,
where $M=1,2,\ldots,2D$, while the worldsheet $\Sigma$ is still
two dimentional. Therefore, $h_{MN}$ and $\eta_{MN}$ are $2D\times2D$
matrices. $g_{ij}$ is the spacetime metric. It is important that
this action has an $O\left(D,D\right)$ symmetry: invariant under
rotation $\Omega$ satisfying $\Omega\eta\Omega^{T}=\eta$. Here for
later convenience, we use $X^{M}=\left(X^{i},\tilde{X}_{i}\right)$
in the action, different from the conventions in \cite{Tseytlin:1990nb} .
In this paper, we assume the open string ending on spacefilling branes.
The case for $Dp-$branes can be easily generalized. It is worth noting
that open and closed strings share the same action. The differences
between open and closed strings are determined by the boundary conditions.
In recent works, the double field theory action of closed strings
has been derived from this action \cite{Berman:2007xn}  . The equation
of motion (EOM) is obtained by varying $X^{M}$ 

\begin{equation}
\partial_{1}\left(h_{MN}\partial_{1}X^{N}-\eta_{MN}\partial_{0}X^{N}\right)=0,\label{eq:Original EOM}
\end{equation}

\noindent which leads to

\begin{eqnarray}
g\partial_{1}X-\partial_{0}\tilde{X} & = & f_{1}\left(\tau\right),\label{eq:constraint1}\\
g^{-1}\partial_{1}\tilde{X}-\partial_{0}X & = & f_{2}\left(\tau\right),\label{eq:constraint2}
\end{eqnarray}

\noindent where $f_{1}\left(\tau\right)$ and $f_{2}\left(\tau\right)$
are arbitrary regular functions solely depending on $\tau$. It turns
out that after imposing the self-duality condition $f_{i}=0$, the
action (\ref{eq:Double Sigma Action}) recedes back to the Polyakov
action \cite{Tseytlin:1990nb}. This self-duality condition corresponds
to the strong constraint in the framework of double field theory \cite{Hull:2004in}.
But we will momentarily keep it unfixed and show that, for open strings,
the self-duality condition is a derived consequence under $O(D,D)$
covariant boundary condition but not a premise. The EOM (\ref{eq:Original EOM})
can be put into decoupled form 

\begin{eqnarray}
(\partial_{1}\,^{2}-\partial_{0}\,^{2})X & = & \partial_{0}f_{2}\left(\tau\right),\label{eq:EOM1}\\
(\partial_{1}\,^{2}-\partial_{0}\,^{2})\tilde{X} & = & \partial_{0}f_{1}\left(\tau\right).\label{eq:EOM2}
\end{eqnarray}

\noindent The boundary conditions of open strings are determined by
the EOM 

\begin{equation}
\left.\delta X^{M}\left(h_{MN}\partial_{1}X^{N}-\frac{1}{2}\eta_{MN}\partial_{0}X^{N}\right)\right|_{\partial\Sigma}=0,
\end{equation}

\noindent which can be expanded as

\begin{equation}
\left.\delta X\left(g\partial_{1}X-\frac{1}{2}\partial_{0}\tilde{X}\right)+\delta\tilde{X}\left(g^{-1}\partial_{1}\tilde{X}-\frac{1}{2}\partial_{0}X\right)\right|_{\partial\Sigma}=0,
\end{equation}

\noindent It looks like we have four options for the boundary conditions.
However, $O(D,D)$ covariance excludes two of them and we are only
left with two equally good boundary conditions 

\begin{eqnarray}
\left.\delta X\right|_{\partial\Sigma} & = & \left.\partial_{0}X\right|_{\partial\Sigma}=0,\nonumber \\
\left.g^{-1}\partial_{1}\tilde{X}-\frac{1}{2}\partial_{0}X\right|_{\partial\Sigma} & = & 0\Rightarrow\partial_{1}\tilde{X}|_{\partial\Sigma}=0.
\end{eqnarray}

\noindent or

\begin{eqnarray}
\left.\delta\tilde{X}\right|_{\partial\Sigma} & = & \left.\partial_{0}\tilde{X}\right|_{\partial\Sigma}=0,\nonumber \\
\left.g\partial_{1}X-\frac{1}{2}\partial_{0}\tilde{X}\right|_{\partial\Sigma} & = & 0\Rightarrow\partial_{1}X|_{\partial\Sigma}=0.\label{eq:boundary condition}
\end{eqnarray}

\noindent Apparently, these two sets of boundary conditions agree
with T-duality of open strings: if one boundary condition is Neumann,
its T-dual boundary condition is Dirichlet. Since we eventually want
to compare with the results in \cite{Seiberg:1999vs}, hereafter we
will select the second boundary condition (\ref{eq:boundary condition}).
After applying (\ref{eq:constraint1}) on the boundary, one gets $f_{1}\left(\tau\right)=0$.
On the other hand, we can make a shift $X\to X-\int d\tau f_{2}\left(\tau\right)$
which does not affect the boundary conditions, EOM and the action.
We are therefore free to set $f_{2}(\tau)=0$. The decoupled EOM now
can be casted into 

\begin{eqnarray}
(\partial_{1}\,^{2}-\partial_{0}\,^{2})X & = & 0,\nonumber \\
\left.\partial_{1}X\right|_{\partial\Sigma} & = & 0,
\end{eqnarray}

\noindent and

\begin{eqnarray}
(\partial_{1}\,^{2}-\partial_{0}\,^{2})\tilde{X} & = & 0,\nonumber \\
\left.\partial_{0}\tilde{X}\right|_{\partial\Sigma} & = & 0.
\end{eqnarray}

\noindent From now on, we work with Euclidean signature by setting
$\tau\rightarrow-i\tau$ ($\partial_{0}\rightarrow i\partial_{0}$),
and use complex coordinates: $\partial_{0}=\partial+\bar{\partial}$
and $\partial_{1}=i\left(\partial-\bar{\partial}\right)$. It is easy
to figure out the propagators 

\begin{equation}
\left\langle X^{i}\left(z,\bar{z}\right)X^{j}\left(z^{\prime},\bar{z}^{\prime}\right)\right\rangle =-\alpha'\left(g^{ij}\log\left|z-z^{\prime}\right|+g^{ij}\log\left|z-\bar{z}^{\prime}\right|\right).
\end{equation}

\begin{equation}
\left\langle \tilde{X}_{i}\left(z,\bar{z}\right)\tilde{X}_{j}\left(z^{\prime},\bar{z}^{\prime}\right)\right\rangle =-\alpha'\left(g_{ij}\log\left|z-z^{\prime}\right|-g_{ij}\log\left|z-\bar{z}^{\prime}\right|\right).
\end{equation}
From these two propagators, $X$ and $\tilde{X}$ are both commutative.
Our next step is to calculate the mixed propagators $\left\langle \tilde{X}_{i}\left(z,\bar{z}\right)X^{j}\left(z^{\prime},\bar{z}^{\prime}\right)\right\rangle $
and $\left\langle X^{i}\left(z,\bar{z}\right)\tilde{X}_{j}\left(z^{\prime},\bar{z}^{\prime}\right)\right\rangle $.
From eqn. (\ref{eq:constraint1}) and (\ref{eq:constraint2}) with
$f_{i}=0$, we have 

\begin{equation}
g_{ij}\partial_{1}X^{j}=i\partial_{0}\tilde{X_{i}},\qquad g^{ij}\partial_{1}\tilde{X}_{j}=i\partial_{0}X^{i}.
\end{equation}

\noindent Therefore, the propagators satisfy the following equations

\begin{eqnarray}
g_{\ell i}\left(\partial-\bar{\partial}\right)\left\langle X^{i}\left(z,\bar{z}\right)\tilde{X}_{j}\left(z^{\prime},\bar{z}^{\prime}\right)\right\rangle  & = & \left(\partial+\bar{\partial}\right)\left\langle \tilde{X}_{\ell}\left(z,\bar{z}\right)\tilde{X}_{j}\left(z^{\prime},\bar{z}^{\prime}\right)\right\rangle ,\\
g^{\ell i}\left(\partial-\bar{\partial}\right)\left\langle \tilde{X}_{i}\left(z,\bar{z}\right)X^{j}\left(z^{\prime},\bar{z}^{\prime}\right)\right\rangle  & = & \left(\partial+\bar{\partial}\right)\left\langle X^{\ell}\left(z,\bar{z}\right)X^{j}\left(z^{\prime},\bar{z}^{\prime}\right)\right\rangle ,
\end{eqnarray}

\noindent The solutions of these two equations are 

\begin{eqnarray}
\left\langle X^{i}\left(z,\bar{z}\right)\tilde{X}_{j}\left(z^{\prime},\bar{z}^{\prime}\right)\right\rangle  & = & -\frac{\alpha'}{2}g^{ik}g_{kj}\left(\log\frac{z-z^{\prime}}{\bar{z}-\bar{z}^{\prime}}-\log\frac{z-\bar{z}^{\prime}}{\bar{z}-z^{\prime}}\right),\\
\left\langle \tilde{X}_{i}\left(z,\bar{z}\right)X^{j}\left(z^{\prime},\bar{z}^{\prime}\right)\right\rangle  & = & -\frac{\alpha'}{2}g_{ik}g^{kj}\left(\log\frac{z-z^{\prime}}{\bar{z}-\bar{z}^{\prime}}+\log\frac{z-\bar{z}^{\prime}}{\bar{z}-z^{\prime}}\right).
\end{eqnarray}

\noindent It is easy to check that these propagators also satisfy
the boundary conditions. Considering the propagators on the boundary
$z=\tau$ and $z'=\tau'$, we can get the commutators
\begin{eqnarray}
[X^{i}(\tau),X^{j}(\tau')] & = & [\tilde{X}^{i}(\tau),\tilde{X}^{j}(\tau')]=0,\nonumber \\
{}[\tilde{X}_{i}(\tau),X^{j}(\tau)] & = & i2\pi\alpha'\delta_{i}\,^{j}.\label{eq:cross noncomm}
\end{eqnarray}

\noindent Therefore, we find that $X$ or $\tilde{X}$ alone is commutative,
but they do not commute with each other. This striking feature comes
from the requirements that the double sigma model is $O(D,D)$ invariant
and can reduce back to the Polyakov action. However, this non-commutativity
is hidden from the ordinary picture. Since the action is $O(D,D)$
invariant, we now go to a general phase frame by a pure coordinate
transformation

\begin{equation}
\Omega=\left(\begin{array}{cc}
1 & -B^{ij}\\
0 & 1
\end{array}\right),\label{eq:rotation}
\end{equation}

\noindent where $B^{ij}$ is an antisymmetric tensor. The generalized
metric $h_{MN}$ is then rotated to

\begin{equation}
H_{MN}=\Omega^{T}h_{MN}\Omega=\left(\begin{array}{cc}
g & -gB^{-1}\\
B^{-1}g & g^{-1}-B^{-1}gB^{-1}
\end{array}\right),\label{eq:General H}
\end{equation}

\noindent accompanied by the coordinate transformation

\begin{eqnarray}
X^{i\prime} & = & X^{i}+B^{ij}\tilde{X}_{j},\nonumber \\
\tilde{X}_{j}^{\prime} & = & \tilde{X}_{j}.
\end{eqnarray}

\noindent It is straightforward to calculate the propagators in the
new frame  

\begin{eqnarray}
 &  & \left\langle X^{i\prime}\left(z,\bar{z}\right)X^{j\prime}\left(z^{\prime},\bar{z}^{\prime}\right)\right\rangle \nonumber \\
 & = & \left\langle X^{i}\left(z,\bar{z}\right)+B^{ik}\tilde{X}_{k}\left(z,\bar{z}\right),X^{j}\left(z^{\prime},\bar{z}^{\prime}\right)+B^{j\ell}\tilde{X}_{l}\left(z^{\prime},\bar{z}^{\prime}\right)\right\rangle \nonumber \\
 & = & \left\langle X^{i}\left(z,\bar{z}\right),X^{j}\left(z^{\prime},\bar{z}^{\prime}\right)\right\rangle -\left\langle X^{i}\left(z,\bar{z}\right),\tilde{X}_{\ell}\left(z^{\prime},\bar{z}^{\prime}\right)\right\rangle B^{\ell j}\nonumber \\
 &  & +B^{ik}\left\langle \tilde{X}_{k}\left(z,\bar{z}\right),X^{j}\left(z^{\prime},\bar{z}^{\prime}\right)\right\rangle -B^{ik}\left\langle \tilde{X}_{k}\left(z,\bar{z}\right),\tilde{X}_{\ell}\left(z^{\prime},\bar{z}^{\prime}\right)\right\rangle B^{lj}\nonumber \\
 & = & -\alpha'\left[\left(g^{ij}-B^{ik}g_{k\ell}B^{\ell j}\right)\log\left|z-z^{\prime}\right|+\left(g^{ij}+B^{ik}g_{k\ell}B^{\ell j}\right)\log\left|z-\bar{z}^{\prime}\right|+B^{ij}\left(\log\frac{z-\bar{z}^{\prime}}{\bar{z}-z^{\prime}}\right)\right].\nonumber \\
\label{eq:Non-Propagator}
\end{eqnarray}

\noindent We see that the non-commutativity becomes visible in the
last term. It is of importance to note that this propagator natively
uses the open string metric and parameters. That is why it looks different
from the propagator in \cite{Seiberg:1999vs} where the propagator
was explicitly expressed with the closed string metric and parameters.
Therefore, in order to make comparison, we need to rotate the parameters
back to the closed string parameters. Rotating $H_{MN}$ back to the
closed string parameters $\hat{g}$ and $\hat{B}$ is achieved by
the transformation

\begin{equation}
\eta\left(\begin{array}{cc}
\hat{g}^{-1} & -\hat{g}^{-1}\hat{B}\\
\hat{B}\hat{g}^{-1} & \hat{g}-\hat{B}\hat{g}^{-1}\hat{B}
\end{array}\right)\eta=\left(\begin{array}{cc}
\hat{g}-\hat{B}\hat{g}^{-1}\hat{B} & \hat{B}\hat{g}^{-1}\\
-\hat{g}^{-1}\hat{B} & \hat{g}^{-1}
\end{array}\right)=\left(\begin{array}{cc}
g & -gB^{-1}\\
B^{-1}g & g^{-1}-B^{-1}gB^{-1}
\end{array}\right).\label{eq:HH}
\end{equation}

\noindent It gives us the following relations

\begin{eqnarray}
g_{ij} & = & \left(\hat{g}-\hat{B}\hat{g}^{-1}\hat{B}\right)_{ij},\nonumber \\
B^{ij} & = & -\left(\frac{1}{\hat{g}+\hat{B}}\hat{B}\frac{1}{\hat{g}-\hat{B}}\right)^{ij},\nonumber \\
\hat{g}^{ij} & = & \left(g^{-1}-B^{-1}gB^{-1}\right)^{ij}\nonumber \\
\hat{B}^{ij} & = & \left(B^{-1}-g^{-1}Bg^{-1}\right)^{ij}\label{eq:transformations}
\end{eqnarray}

\noindent which are precisely the transformations between closed and
open parameters. Therefore, the propagator (\ref{eq:Non-Propagator})
can be rewritten with the closed string (hat) parameters as

\begin{eqnarray}
\left\langle X^{i\prime}\left(z,\bar{z}\right)X^{j\prime}\left(z^{\prime},\bar{z}^{\prime}\right)\right\rangle  & = & -\alpha'\bigg[\hat{g}^{ij}\log\left|z-z^{\prime}\right|-\hat{g}^{ij}\log\left|z-\bar{z}^{\prime}\right|\label{eq:F propagator}\\
 &  & \left.+\left(\frac{1}{\hat{g}+\hat{B}}\hat{g}\frac{1}{\hat{g}-\hat{B}}\right)^{ij}\log\left|z-\bar{z}^{\prime}\right|^{2}+\left(-\frac{1}{\hat{g}+\hat{B}}\hat{B}\frac{1}{\hat{g}-\hat{B}}\right)^{ij}\left(\log\frac{z-\bar{z}^{\prime}}{\bar{z}-z^{\prime}}\right)\right],\nonumber 
\end{eqnarray}

\noindent which is completely identical to the Seiberg-Witten result
in \cite{Seiberg:1999vs}. It is of help to give the other three propagators
with the closed string (hat) parameters

\begin{eqnarray}
\left\langle \tilde{X}_{i}^{\prime}\left(z,\bar{z}\right)\tilde{X}_{j}^{\prime}\left(z^{\prime},\bar{z}^{\prime}\right)\right\rangle  & = & -\alpha'\left[\left(\hat{g}-\hat{B}\hat{g}^{-1}\hat{B}\right)_{ij}\log\left|z-z^{\prime}\right|-\left(\hat{g}-\hat{B}\hat{g}^{-1}\hat{B}\right)_{ij}\log\left|z-\bar{z}^{\prime}\right|\right],\label{eq:dual propagator}\\
\left\langle X^{i\prime}\left(z,\bar{z}\right)\tilde{X}_{j}^{\prime}\left(z^{\prime},\bar{z}^{\prime}\right)\right\rangle  & = & -\frac{\alpha'}{2}\delta_{\; j}^{i}\left(\log\frac{z-z^{\prime}}{\bar{z}-\bar{z}^{\prime}}-\log\frac{z-\bar{z}^{\prime}}{\bar{z}-z^{\prime}}\right)\nonumber \\
 &  & -\alpha'\left(\frac{1}{\hat{g}+\hat{B}}\hat{B}\frac{1}{\hat{g}-\hat{B}}\right)_{\; j}^{i}\left(\log\left|z-z^{\prime}\right|-\log\left|z-\bar{z}^{\prime}\right|\right),\\
\left\langle \tilde{X}_{i}^{\prime}\left(z,\bar{z}\right)X^{j\prime}\left(z^{\prime},\bar{z}^{\prime}\right)\right\rangle  & = & -\frac{\alpha'}{2}\delta_{i}^{\; j}\left(\log\frac{z-z^{\prime}}{\bar{z}-\bar{z}^{\prime}}+\log\frac{z-\bar{z}^{\prime}}{\bar{z}-z^{\prime}}\right)\nonumber \\
 &  & +\alpha'\left(\frac{1}{\hat{g}+\hat{B}}\hat{B}\frac{1}{\hat{g}-\hat{B}}\right)_{i}^{\; j}\left(\log\left|z-z^{\prime}\right|-\log\left|z-\bar{z}^{\prime}\right|\right).
\end{eqnarray}

\noindent We can easily see the commutators 

\noindent 
\begin{eqnarray*}
[X^{i\prime}\left(\tau\right),X^{j\prime}\left(\tau\right)] & = & i\pi\alpha'\left(-\frac{1}{\hat{g}+\hat{B}}\hat{B}\frac{1}{\hat{g}-\hat{B}}\right)^{ij}\\
{}[\tilde{X}'_{i}(\tau),\tilde{X}_{j}'(\tau)] & = & 0\\
{}[\tilde{X}'_{i}(\tau),X^{j\prime}\left(\tau\right)] & = & i2\pi\alpha'\delta_{i}\,^{j}.
\end{eqnarray*}
To see how the Seiberg-Witten map arises, we eliminate $\tilde{X}$
from the double sigma action (\ref{eq:Double Sigma Action}) in general
frames. Using the EOM of $\tilde{X}$ and the boundary condition $\delta\tilde{X}|_{\partial\Sigma}=0,$
after a bit calculation, we get the action with Euclidean signature

\begin{equation}
S=\frac{1}{2}\int_{\Sigma}\,\partial_{a}X'\,\left(\frac{1}{g^{-1}-B^{-1}}g^{-1}\frac{1}{g^{-1}+B^{-1}}\right)\,\partial^{a}X'-2i\partial_{0}X\left(-\frac{1}{g^{-1}-B^{-1}}B^{-1}\frac{1}{g^{-1}-B^{-1}}\right)\partial_{1}X,\label{eq:reduced action}
\end{equation}

\noindent which confirms eqn. (\ref{eq:HH}). It is readily to obtain
the DBI of this action

\begin{eqnarray*}
S_{DBI} & = & \frac{1}{g_{s}}\int d^{D}x\sqrt{\det\left(\frac{1}{g^{-1}+B^{-1}}+F\right)}\\
 & = & \frac{1}{g_{s}}\int d^{D}x\sqrt{\hat{g}+\hat{B}+F}\\
 & = & \frac{1}{G_{s}}\int d^{D}x\sqrt{\det\left(g+F^{*}\right)},
\end{eqnarray*}

\noindent with

\begin{equation}
F^{*}=\frac{1}{1+FB^{-1}}F,\qquad\qquad G_{s}\equiv g_{s}\sqrt{\det\left(1+B^{-1}g\right)},\label{eq:simple map}
\end{equation}

\noindent where $F^{*}$ is the non-commutative gauge field strength
and $F$ is the commutative one. It is identical to that of Seiberg-Witten,
but the physical interpretation is now changed: $B^{-1}$ is an $O(D,D)$
group parameter. It is now clear that in the original work of Seiberg-Witten,
only one set of coordinates is visible. There is no non-commutativity
at the very beginning when $\hat{B}=0$, since the mixed non-commutative
propagators $\left\langle \tilde{X}{}^{i}\left(z,\bar{z}\right)X_{j}\left(z^{\prime},\bar{z}^{\prime}\right)\right\rangle $
is hidden. Non-commutativity emerges upon turning on a boundary term.
However, when we begin with the double sigma model, non-commutativity
is an intrinsic property and one can freely transform a commutative
sector to a non-commutative one or vice versa. The Seiberg-Witten
map is really just a symmetry. More importantly, the metric or parameters
in the double sigma model are natively determined to be open or closed
parameters by the action itself upon imposing $O(D,D)$ covariant
boundary conditions as demonstrated in eqn. (\ref{eq:reduced action}). 

\noindent It is interesting to look at the propagator (\ref{eq:dual propagator}),
which is commutative even if we have a non-vanishing $\hat{B}$ field.
This propagator extends the results of Seiberg-Witten map. In an abstract
way, we generalize the open-closed string relation from equations
(\ref{eq:HH}) and (\ref{eq:F propagator}):

\begin{equation}
\left(\begin{array}{cc}
\hat{g}-\hat{B}\hat{g}^{-1}\hat{B} & \hat{B}\hat{g}^{-1}\\
-\hat{g}^{-1}\hat{B} & \hat{g}^{-1}
\end{array}\right)=\left(\begin{array}{cc}
G & G\theta\\
-\theta G & G^{-1}-\theta G\theta
\end{array}\right),
\end{equation}

\noindent where we identify the open string metric $G\equiv g$ and
$\theta\equiv B$. It gives a generalized map between open string
variables $\left(G_{s},G_{ij},\theta_{ij}\right)$ and closed string
variables $\left(g_{s},\hat{g}_{ij},\hat{B}_{ij}\right)$. When $\hat{B}=0$,
we get

\begin{equation}
\left(\begin{array}{cc}
\hat{g} & 0\\
0 & \hat{g}^{-1}
\end{array}\right)=\left(\begin{array}{cc}
G & 0\\
0 & G^{-1}
\end{array}\right),\quad G=g
\end{equation}

\noindent where two sides are all commutative. However, when $\hat{B}\neq0$,
we can also obtain commutativity through

\begin{equation}
\left(\begin{array}{cc}
1 & 0\\
\theta & 1
\end{array}\right)\left(\begin{array}{cc}
G & G\theta\\
-\theta G & G^{-1}-\theta G\theta
\end{array}\right)\left(\begin{array}{cc}
1 & -\theta\\
0 & 1
\end{array}\right)=\left(\begin{array}{cc}
G & 0\\
0 & G^{-1}
\end{array}\right),\quad G=\hat{g}-\hat{B}\hat{g}^{-1}\hat{B},
\end{equation}

\noindent which corresponds to the propagator (\ref{eq:dual propagator}).
Therefore, when $B\neq0$, there also exists commutativity.

\section{The general descriptions of Seiberg-Witten map }

From eqn. (\ref{eq:simple map}) and (\ref{eq:transformations}),
the {*} product of the non-commutative Yang-Mills theory $F^{*}$
is defined by a definite quantity: $B^{-1}=-(\hat{g}+\hat{B})^{-1}\hat{B}(\hat{g}-\hat{B})^{-1}$.
In \cite{Seiberg:1999vs}, Seiberg and Witten proposed that there
exist more general descriptions of the map, with an arbitrary parameter
$\theta$ but not just $B^{-1}$ and the $\theta$ dependence of the
effective action is completely captured by replacing

\begin{equation}
F^{*}\to F^{*}+\Phi,
\end{equation}
where $\Phi$ is some two-form depending on $\hat{B}$, $\hat{g}$
and $\theta$, called the \textit{description parameter}, determined
by 

\begin{eqnarray}
\frac{1}{g+\Phi} & = & -\theta+\frac{1}{\hat{g}+\hat{B}}\label{eq:gen descr condition-1}\\
G_{s} & = & g_{s}\left(\frac{\det\left(g+\Phi\right)}{\det\left(\hat{g}+\hat{B}\right)}\right)^{\frac{1}{2}}=g_{s}\frac{1}{\det\left[\left(\frac{1}{\hat{g}+\hat{B}}-\theta\right)\left(\hat{g}+\hat{B}\right)\right]^{\frac{1}{2}}}\label{eq:gen descr condition-2}
\end{eqnarray}
with the non-commutative effective action 

\begin{equation}
\mathcal{L}_{DBI}^{*}=\frac{1}{G_{s}}\sqrt{\det\left(G+F^{*}+\Phi\right)}
\end{equation}
To see what we can get from $O(D,D)$ on general descriptions, we
make a second rotation following that of (\ref{eq:rotation}) 

\begin{equation}
\Omega'=\left(\begin{array}{cc}
1 & -B^{-1}\\
0 & 1
\end{array}\right)\left(\begin{array}{cc}
1 & 0\\
-C & 1
\end{array}\right)=\left(\begin{array}{cc}
1+B^{-1}C & -B^{-1}\\
-C & 1
\end{array}\right),
\end{equation}
where $B$ and $C$ are two-forms. The generalized metric under this
rotation is

\begin{equation}
H'_{MN}=\left(\begin{array}{cc}
g+CB^{-1}g+gB^{-1}C-Cg^{-1}C+CB^{-1}gB^{-1}C & -gB^{-1}+Cg^{-1}-CB^{-1}gB^{-1}\\
B^{-1}g-g^{-1}C+B^{-1}gB^{-1}C & g^{-1}-B^{-1}gB^{-1}
\end{array}\right).
\end{equation}
The corresponding coordinate transformations are 

\begin{eqnarray*}
X' & = & X+B^{-1}\tilde{X}\\
\tilde{X}' & = & CX+(1+CB^{-1})\tilde{X}
\end{eqnarray*}
It is easy to imagine that with these coordinate transformations,
both $X'$ and $\tilde{X'}$ are non-commutative. Applying (\ref{eq:HH}),
we identify

\begin{eqnarray}
\hat{g}^{-1} & = & g^{-1}-B^{-1}gB^{-1}\nonumber \\
\hat{B} & = & C-\frac{1}{g^{-1}+B^{-1}}B^{-1}\frac{1}{g^{-1}-B^{-1}}\nonumber \\
\hat{g}+\hat{B} & = & C+\frac{1}{g^{-1}+B^{-1}}\label{eq:gen trans}
\end{eqnarray}
Then we have the commutative DBI

\begin{eqnarray*}
S_{DBI} & = & \frac{1}{g_{s}}\int d^{D}x\sqrt{\det(\hat{g}+\hat{B}+F)}\\
 & = & \frac{1}{g_{s}}\int d^{D}x\sqrt{\det\left(\frac{1}{g^{-1}+B^{-1}}+F+C\right)}\\
\end{eqnarray*}
After a bit calculations, one can prove that the non-commutative description
is

\begin{equation}
S_{DBI}=\frac{1}{G_{s}}\int d^{D}x\sqrt{\det\left(g+F^{*}+\Phi\right)},\label{eq:Non-com DBI}
\end{equation}
with $G_{s}$ defined by (\ref{eq:gen descr condition-2}) and the
constraint for $\Phi$ and $\theta$ 
\begin{equation}
\frac{1}{g+\Phi}+\theta=\frac{1}{C+\frac{1}{g^{-1}+B^{-1}}}=\frac{1}{\hat{g}+\hat{B}},\label{eq:gen constraint}
\end{equation}
and 

\begin{equation}
F^{*}=\frac{1}{1+F\theta}F.
\end{equation}
A remarkable observation is that from (\ref{eq:gen trans}), for fixed
closed string $\hat{g}$ and $\hat{B}$, the open string parameters
$g$ and $B$ are still free to vary provided $C$ varying accordingly.
Referring to (\ref{eq:gen constraint}), $\Phi$ is unnecessary to
keep $\theta$ varying and we can therefore set $\Phi=0$ in (\ref{eq:Non-com DBI})
and (\ref{eq:gen constraint}). Therefore, $C$ actually plays the
role of the description parameter and in the non-commutative DBI,
its effect is incorporated into the variation of the open string metric
$g$. This tells us that one does not need to introduce an independent
field and the $O(D,D)$ symmetry already has this ingredient.

\bigskip{}

In summary, we start from the double sigma model and found that for
open strings, the theory is intrinsically non-commutative. In the
simplest phase state, the ordinarily visible sector is commutative
and the non-commutativity is hidden. In a general frame, the non-commutativity
arises from $O(D,D)$ rotations. The visible sector completely agrees
with previous results. We showed that the parameters of the double
sigma model are determined to be open or closed by the boundary conditions.
The open and closed parameters are related by an $O(D,D)$ transformation.
We further exhibited that as $\hat{B}\not=0$, besides the non-commutative
theory, there is another commutative one. Our results demonstrated
that the Seiberg-Witten map is actually a subset of $O(D,D)$ symmetry.
Finally, we explored double rotations and found that the general descriptions
of Seiberg-Witten map is also naturally derived by a group parameter.
In this scenario, all the sectors are non-commutative. It is quite
interesting that there are a lot of similarities between our results
and the non-commutativity from closed strings \cite{Andriot:2011uh}.
These similarities cannot be accident and need more explorations in
the future works.

\vspace{5mm}

\noindent {\bf Acknowledgements} 
We would like to acknowledge illuminating discussions with Zheng Sun, Yan He and Bo Feng. 
This work is supported in part by the NSFC (Grant No. 11175039 and 11375121 ) and SiChuan Province Science Foundation for Youths (Grant No. 2012JQ0039). 

\end{document}